\documentclass[a4paper,10pt]{article}
\usepackage[utf8]{inputenc}

\usepackage[numbers,sort&compress]{natbib}
\usepackage{graphicx}
\usepackage{amsmath}
\usepackage{authblk} 

\usepackage{hyperref} 

\usepackage[margin=1in]{geometry} 

\begin{document}

\begin{minipage}[t]{1\textwidth}
\begin{flushleft} \Large
\textbf{Systematic study of nonmagnetic resistance changes due to electrical pulsing in single metal layers and metal/antiferromagnet bilayers}
\\[1cm] 
\large 
B. J. Jacot$^{1}$, G. Krishnaswamy$^{1}$, G. Sala$^{1}$, C. O. Avci$^{1}$, S. Vélez$^{1}$, P. Gambardella$^{1}$, C.-H. Lambert$^{1}$ \\
$^{1}$Department of Materials, ETH Zurich, 8093 Zurich, Switzerland \\
email: \href{mailto:benjamin.jacot@mat.ethz.ch}{benjamin.jacot@mat.ethz.ch} \\
email: \href{mailto:charles-henri.lambert@mat.ethz.ch}{charles-henri.lambert@mat.ethz.ch}
\date{}
\end{flushleft}
\end{minipage}

\section*{\label{sec:abstract} Abstract} Intense current pulses are often required to operate microelectronic and spintronic devices. Notably, strong current pulses have been shown to induce magnetoresistance changes attributed to domain reorientation in antiferromagnet/heavy metal bilayers and non-centrosymmetric antiferromagnets. In such cases, nonmagnetic resistivity changes may dominate over signatures of antiferromagnetic switching. We report systematic measurements of the current-induced changes of the transverse and longitudinal resistance of Pt and Pt/NiO layers deposited on insulating substrates, namely Si/SiO$_2$, Si/Si$_3$N$_4$, and Al$_2$O$_3$. We identify the range of pulse amplitude and length that can be used without affecting the resistance and show that it increases with the device size and thermal diffusivity of the substrate. No significant difference is observed in the resistive response of Pt and NiO/Pt devices, thus precluding evidence on the switching of antiferromagnetic domains in NiO. The variation of the transverse resistance is associated to a thermally-activated process in Pt that decays following a double exponential law with characteristic timescales of a few minutes to hours. We use a Wheatstone bridge model to discriminate between positive and negative resistance changes, highlighting competing annealing and electromigration effects. Depending on the training of the devices, the transverse resistance can either increase or decrease between current pulses. Further, we elucidate the origin of the nonmonotonic resistance baseline, which we attribute to training effects combined with the asymmetric distribution of the current. These results provide insight into the origin of current-induced resistance changes in metal layers and a guide to minimize nonmagnetic artifacts in switching experiments of antiferromagnets.

\clearpage

\section{\label{sec:intro}Introduction}

The compensated magnetic lattice and the strong exchange coupling between antiparallel magnetic moments make antiferromagnets (AFMs) appealing for data storage applications \cite{A036-Olejnik2017, A002-Jungwirth2016}. The absence of a net magnetization limits the cross-talk between devices in densely-packed layouts and minimizes the influence of external magnetic fields on the magnetic order parameter. Moreover, the dynamic resonant modes associated with the staggered antiferromagnetic structure are orders of magnitude higher in frequency compared to ferromagnets \cite{A178-Keffer1951, A203-Fiebig2008}, thus opening unique prospects for terahertz spintronics \cite{A025-Baltz2018, A029-kampfrath2018}. For a long time, however, controlling the orientation of the magnetic moments in AFMs by means of scalable and integration-friendly methods was deemed to be impossible. Recent experiments provide a solution to this problem, as it was shown that current injection in non-centrosymmetric metallic AFMs \cite{A014-Wadley2016, A181-Wadley2018, A182-Zhou2018, A029-kampfrath2018, A039-Meinert2018, A045-Bodnar2018, A046-Chen2019, A107-Nair2019, A164-Dunz2020, A200-Omari2020, A211-Tsai2020} as well as in heavy metal layers deposited on insulating AFMs \cite{A004-Moriyama2018, A003-Chen2018, A070-Baldrati2019, A115-Gray2019, A163-Zhang2019, A210-Cheng2020} lead to the in-plane rotation of the Néel vector due to the current-induced spin-orbit torques \cite{R002-Manchon2019}. Reading of the Néel vector reorientation in metallic AFMs and AFM/heavy metal bilayers can then be performed by measuring the anisotropic magnetoresistance of the AFM and spin Hall magnetoresistance of the heavy metal, respectively. Together, these effects allow for all-electrical manipulation and detection of antiferromagnetic order.

The electrical switching of AFMs requires a sufficient torque to reorient the magnetic moments, which in turn requires a high current density. In addition to the torque, Joule heating plays a critical role in the switching by assisting the magnetic moments in overcoming the magnetic anisotropy energy barrier between two different easy axis orientations \cite{A039-Meinert2018, A073-Wornle2019}. Intense currents, however, can also produce irreversible changes of the materials due to thermal annealing \cite{A204-Schmid2008} and electromigration \cite{A205-Hi1989}. Although the switching of the Néel vector in AFM/heavy metal bilayers such as NiO/Pt has been confirmed by imaging techniques, including x-ray photoemission electron microscopy \cite{A004-Moriyama2018, A070-Baldrati2019}, birefringence \cite{A317-Schreiber2020} and spin Seebeck microscopy \cite{A115-Gray2019}, transport measurements on single-layer Nb and Pt thin films have shown nonmagnetic resistive signatures very similar to the ones attributed to magnetoresistance changes \cite{A072-Chiang2019, A069-Wagner2019, A138-Churikova2020}. Analogous resistive signatures in AFM/Pt bilayers are considered as a proof of current-induced switching in an increasing variety of AFMs, such as hematite \cite{A163-Zhang2019, A210-Cheng2020}, Mn$_2$Au \cite{A041-Zhou2019}, PtMn \cite{A206-Shi2020}, MnN \cite{A164-Dunz2020}, Mn$_3$Sn \cite{A211-Tsai2020} and synthetic antiferromagnets \cite{A048-Moriyama2018}. Given the multiple factors that influence the resistivity of these bilayers, it is important to investigate the pulsing conditions and device characteristics that lead to the appearance of nonmagnetic resistance changes in metallic systems, containing AFMs or not. 

In this study, we perform a systematic characterization of the current-induced resistance changes in Pt and NiO/Pt cross-shaped devices deposited on different substrates. We report measurements of the transverse and longitudinal resistance as a function of the amplitude and length of the current pulses, device size, and training history. In agreement with the nonmagnetic resistive effects revealed by previous studies \cite{A072-Chiang2019, A069-Wagner2019, A138-Churikova2020}, we show that both gradual and step-like resistance changes can be induced in Pt depending on the training of the devices and independently of the presence of NiO. Therefore, the resistance variations in NiO/Pt cannot be considered as signatures of domain switching in NiO. Using a simple Wheatstone bridge model, we identify competing effects that either decrease or increase the local resistivity, which we attribute to current-induced annealing and electromigration, respectively. Moreover, we show that the resistance changes depend strongly on the pulse length and training history of the devices as well as on the device size, which results in a broad variety of resistive signals.

This paper is organized as follows: Section~\ref{sec:pulsing-and-reading-scheme} describes the pulsing and read-out schemes commonly employed in AFM devices and in this work. Sections~\ref{sec:wheatstone-bridge} and \ref{sec:experiment} introduce the Wheatstone bridge model used to interpret the nonmagnetic resistance changes and the experimental systems, respectively. Sections~\ref{sec:results-rxy-pulse-amplitude-const-pulse-length} and \ref{sec:results-temporal-relaxation} present the behavior of the transverse resistance upon application of pulses of increasing amplitude and the relaxation process after excitation. Sections~\ref{sec:results-rxy-pulse-amplitude-pulse-length} and \ref{sec:results-influence-substrate} report the device response as a function of the pulse length and the substrate thermal diffusivity. Section~\ref{sec:results-size-effects} and \ref{sec:results-training-effects} focus on the influence of the device size in the presence of artifacts and how asymmetric responses between two pulsing directions affect the behavior of the transverse resistance.

\section{\label{sec:pulsing-and-reading-scheme} Electrical writing and reading scheme of antiferromagnets}

In metallic AFMs with non-centrosymmetric crystal structure, switching is usually ascribed to the inverse spin galvanic effect \cite{A075-Zelezny2014}. Owing to spin-orbit coupling and the lack of inversion symmetry, an electric current flowing in these AFMs generates a non-equilibrium spin polarization of opposite sign on each magnetic sublattice. The resulting staggered torque tends to align the Néel vector perpendicular to the current. Reversible switching can be achieved by pulsing along two orthogonal directions, thus inducing a rotation of $\pm 90^\circ$ of the Néel vector. This manipulation of the magnetic orientation can then be electrically detected via a transverse magnetoresistance measurement \cite{A014-Wadley2016}. The inverse spin galvanic effect was initially predicted for Mn$_2$Au \cite{A075-Zelezny2014} and experimentally confirmed in CuMnAs \cite{A014-Wadley2016, A181-Wadley2018, A029-kampfrath2018, A200-Omari2020}, Mn$_2$Au \cite{A182-Zhou2018, A039-Meinert2018, A045-Bodnar2018, A046-Chen2019}, and in magnetically intercalated transition metal dichalcogenides \cite{A107-Nair2019}.

In insulating AFMs, the switching process relies on the spin-current transferred from an adjacent metallic layer with strong spin-orbit coupling. The latter scheme is similar to spin-orbit torque switching demonstrated for metallic ferromagnet \cite{A207-Miron2011} and insulating ferrimagnets \cite{A208-Avci2017, R002-Manchon2019}. Several mechanisms have been proposed in order to explain the perpendicular switching of AFMs due to a non-staggered spin torque. In NiO(001)/Pt bilayers, it was suggested that the damping-like spin-orbit torque originating from the current-induced spin accumulation at the Pt interface induces coherent switching of antiferromagnetic domains by orienting the Néel vector towards the writing current \cite{A003-Chen2018}. Alternatively, the field-like spin-orbit torque acting on the uncompensated interfacial spins of NiO could drive a coherent rotation of the spins perpendicular to the writing current in Pt/NiO(111)/Pt trilayers \cite{A004-Moriyama2018}. Finally, mechanisms based on the action of the spin current on antiferromagnetic domain walls have been suggested in NiO/Pt \cite{A082-Tveten2014, A070-Baldrati2019, A115-Gray2019}. A spin current would directly act on the magnetic moments in the domain walls, driving walls with opposite chirality in opposite directions \cite{A217-Shiino2016}. As this process does not discriminate between the expansion or contraction of domains, an additional degeneracy-breaking mechanism is required in order to achieve net switching. The latter is provided by a translational ponderomotive force due to the damping-like torque, which favors the expansion of domains with Néel vector parallel to the current \cite{A070-Baldrati2019}. Recently, an additional current-induced mechanism that does not involve SOT switching was proposed for a hematite/Pt bilayer deposited on Al$_2$O$_3$ \cite{A163-Zhang2019} as well as for NiO/Pt deposited on MgO \cite{A274-Meer2020}, wherein the current-induced Joule heating produces thermal expansion of the substrate and results in mechanical stress that couples to the magnetic order of the AFM through its high magnetostrictive coefficient.

All these switching mechanisms are based on a deterministic writing scheme that aligns the Néel vector either parallel or perpendicular to the writing current, depending on the driving torque being damping-like or field-like, respectively, or magnetostriction. Following one or more writing pulses, an electrical read-out is performed by measuring the variation of the transverse resistance. In metallic AFMs the read-out signal is the transverse component of the anisotropic magnetoresistance --- also called the planar Hall effect --- which is sensitive to the in-plane orientation of the Néel vector relative to the current \cite{A093-Nunez2006,A014-Wadley2016}. In insulating AFMs adjacent to a heavy metal, the signal relates to the transverse component of the spin-Hall magnetoresistance, which is also sensitive to the in-plane orientation of the spins relative to the current direction \cite{A028-Chen2013, A034-Hou2017, A054-Fisher2018, A013-Baldrati2018}. The anisotropic magnetoresistance and spin-Hall magnetoresistance have the common property of being quadratic in the magnetization, which means that they are invariant upon 180$^\circ$ reversal of the Néel vector. Both effects are extremal when the Néel vector sets at $+45^\circ$ ($-45^\circ$) or, equivalently, at $-135^\circ$ ($+135^\circ$) to the sensing current. 

To summarize, the writing process described above accounts for 90$^\circ$ switching of the domains either towards or perpendicular to the writing current, and the read-out signal amplitude is maximal when the sensing current flows at an angle of $\pm 45^\circ$ from the writing line. Hence, the common device geometry is a Hall cross with either 4 or 8 symmetrical arms. In the 4 arms geometry, the set writing pulse, henceforth called P1, is applied along one diagonal of the cross, whereas the reset writing pulse, henceforth called P2, is applied along the other diagonal and orthogonal to P1. During read-out, a small sense current is injected between two opposite arms of the cross and the transverse resistance is measured by probing the Hall voltage between the two arms perpendicular to the sensing current, as illustrated in Figure~\ref{fig: NiO-Pt-comparison}(a). In this work, we consider only the 4 arms geometry, even though similar results are obtained on 8-arms crosses.

\section{\label{sec:wheatstone-bridge} Wheatstone bridge model of a Hall cross}

\begin{figure*}[]
 \centering
 \includegraphics[width=0.9\linewidth]{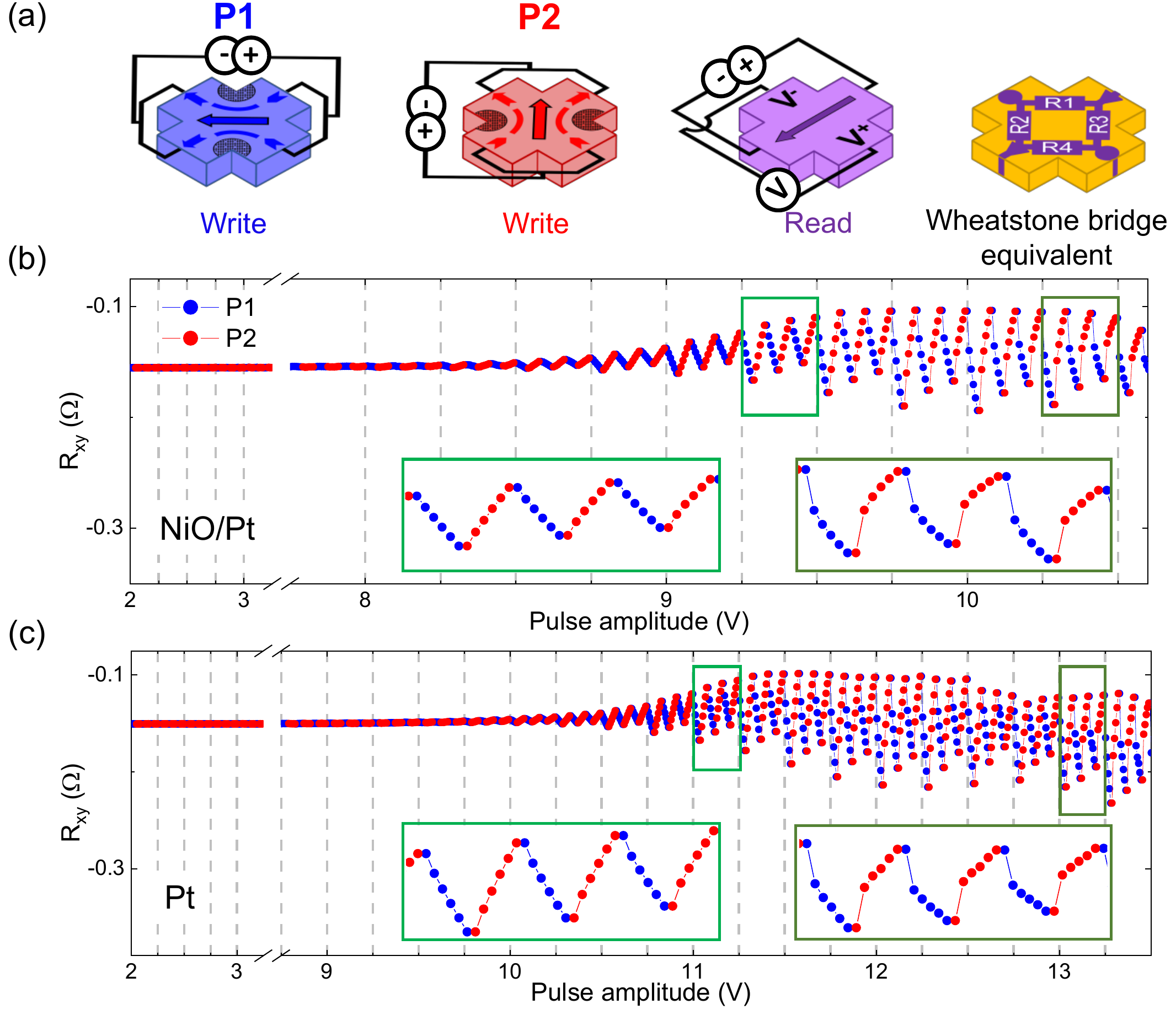}
 \caption{\footnotesize (a) Illustrative diagrams of the pulsing and reading schemes and Wheatstone bridge model of a Hall cross with the corresponding resistances ${R_1, R_2, R_3}$ and ${R_4}$. (b) ${R_{xy}}$ as a function of writing pulse amplitude measured in Al$_2$O$_3$/NiO/Pt and (c) Al$_2$O$_3$/Pt 10 $\mu$m-wide Hall crosses. The pulse amplitude is increased in steps of 0.25~V in correspondence of the gray dashed lines. For each step, a sequence of 3 repeats, made of 5 consecutive P1 pulses and 5 consecutive P2 pulses of length $\tau = 1$~ms is applied. The transverse resistance measured after each pulse is plotted as a blue or red dot for P1 and P2 pulses, respectively. The insets show the saw tooth (left) and step-like (right) resistance changes.}
 \label{fig: NiO-Pt-comparison}
\end{figure*}

According to the considerations presented above, the change of transverse voltage, $\Delta {V_{xy}}$, due to the switching of antiferromagnetic domains should have opposite sign following the writing pulses P1 and P2. Furthermore, if the switching volume is reversible, the signal should be symmetric in amplitude for P1 and P2 pulses. As shown in this study, however, similar transverse voltage variations to the ones expected for the signature of AFM switching can be observed in Hall crosses due to purely nonmagnetic effects. In order to understand the effects of the pulses onto a nonmagnetic resistive device, it is convenient to model the Hall cross as a Wheatstone bridge. We consider a division of the Hall cross into four quadrants of equal size and corresponding resistances ${R_1}$, ${R_2}$, ${R_3}$ and ${R_4}$, as depicted in Fig.~\ref{fig: NiO-Pt-comparison}(a). During the pulsing, the current density is higher around the corners due to the current crowding effect\cite{A139-Hagedorn1963}. The intense heat generated by the current can affect the metallic structure, which in return would change the resistance of the quadrants: P1 would modify mostly ${R_1}$ and ${R_4}$ whereas P2 would modify mostly ${R_2}$ and ${R_3}$. As the sensing current is deflected proportionally to the local resistance, a net voltage arises at the junction of the quadrants, as expected for an unbalanced Wheatstone bridge. For the sensing configuration illustrated in the rightmost panel of Fig.~\ref{fig: NiO-Pt-comparison}(a), the transverse voltage is ${V_{xy} = (R_1/(R_1+R_2) - R_3/(R_3+R_4)) \cdot V_s} $, where ${V_s}$ is the voltage applied by the sensing current source. The resistances of the arms can be discarded in this model as the arms through which the sensing current flows do not influence the transverse voltage and there is no current passing through the voltage arms.

The Wheatstone bridge model highlights two important characteristics of the measurement scheme of AFMs in cross-shape devices. Firstly, ${V_{xy} = 0}$ when ${R_1\cdot R_4= R_2 \cdot R_3}$. Secondly and most importantly, the change of transverse voltage consecutive to a pulse, $\Delta V_{xy}$, caused by the variation of ${R_1}$ and ${R_4}$ has opposite sign than the one caused by the variation of ${R_2}$ and ${R_3}$. Specifically, $\Delta {V_{xy}}$ is negative for a decrease of ${R_1}$ or ${R_4}$, but positive for a decrease of ${R_2}$ or ${R_3}$. Moreover, $\Delta {V_{xy}}$ is positive for a increase of ${R_1}$ or ${R_4}$, but negative for a increase of ${R_2}$ or ${R_3}$. This is a crucial point as it shows that a negative $\Delta {V_{xy}}$ after P1 means that the resistance at the corners has decreased relative to the unpulsed state, whereas a positive $\Delta {V_{xy}}$ after P1 means that the resistance has increased. Thus the sign of $\Delta {V_{xy}}$ allows for discriminating current-induced annealing effects, which tend to reduce the resistance of metal layers, from temporary thermoresistive effects and permanent electromigration effects, which tend to increase the resistance.

\section{\label{sec:experiment} Device fabrication and experimental setup}

We focus on two different sets of samples: a series of Pt(5nm) reference layers and a series of NiO(10nm)/Pt(5nm) bilayers deposited on three different substrates. The numbers between brackets indicate the thickness of each layer. NiO was grown at 550~$^\circ$C, followed by Pt deposition at room temperature after cooling down under vacuum. These layers were sputter-deposited on three different substrates, Si/SiO$_2$(500nm), Si/Si$_3$N$_4$(400nm) and sapphire (Al$_2$O$_3$), at the same time in order to minimize thickness variations. In these conditions, we obtain a typical roughness of less than 1 nm; all layers are preferentially oriented along the (111)-growth direction. X-ray diffraction measurements indicate a high crystalline epitaxial quality of the NiO layers grown on Al$_2$O$_3$, whereas no diffraction peaks are observed for NiO deposited on Si/SiO$_2$(500nm) and Si/Si$_3$N$_4$(400nm) for such thickness. A separate batch deposited on Al$_2$O$_3$ including a Fe layer on top of NiO confirmed the presence of antiferromagnetic order by means of a reversible exchange bias upon cooling in opposite magnetic fields. All layers were subsequently patterned using reactive ion etching in single and double Hall crosses of different width, from 2.5 $\mu$m to 12.5 $\mu$m in steps of 0.5 $\mu$m. The resistivity of each sample was measured with 4 points probes in double Hall crosses, giving ${\rho_\text{{SiO/Pt}}} = 3.20\cdot 10^{-7} $~$\Omega\,$m, ${\rho_\text{{SiO/NiO/Pt}}} = 3.30\cdot 10^{-7} $~$\Omega\,$m, ${\rho_\text{{SiN/Pt}}} = 3.12\cdot 10^{-7} $~$\Omega\,$m, ${\rho_\text{{SiN/NiO/Pt}}} = 3.20\cdot 10^{-7} $~$\Omega\,$m, ${\rho_{{Al_2O_3/Pt}}} = 2.38\cdot 10^{-7} $~$\Omega\,$m, ${\rho_{{Al_2O_3/NiO/Pt}}} = 2.17\cdot 10^{-7} $~$\Omega\,$m. Although the NiO layer does not significantly affect the Pt resistivity, the high crystalline quality of the NiO and Pt films deposited on Al$_2$O$_3$ leads to a lower resistivity than the Si-based substrates. These resistivities are comparable to those obtained for Pt using similar fabrication methods and deposition conditions \cite{A006-Stamm2017, A191-Agustsson2008, A193-Schmid2008}.

The writing pulses were provided by a voltage source that generates rectangular pulses of variable width (50~ns - 1~ms) and amplitude (2~V - 40~V). For instance, a pulse of 10~V corresponds to a current density along the diagonal ${j_w = V/(R_p \, t \, \sqrt{2} w)} = 0.71\cdot10 ^{12}$~A/m$^2$, where ${V}$ is the pulse amplitude, ${R_p} = 200$~$\Omega$ is the pulse path resistance measured at the output terminals (typical value), ${t} = 5$~nm is the film thickness and ${\sqrt{2}w} = \sqrt{2}\cdot10$~$\mu$m is the diagonal of the cross. For reading the transverse voltage, a current source provided an alternating current with peak-to-peak amplitude of 0.5~mA at a frequency of 10~Hz, corresponding to a current density in the sensing arms of $10^{10}$~A/m$^2$ in the same device. The alternating current substantially lowers the noise limit using a pseudo-lock-in detection scheme. The transverse voltage was measured 2~s after pulsing to allow the device to cool down using an integration time of 0.5~s. A relay matrix alternates between the writing (P1 and P2) and the reading configurations. Those relays are essential to control the current path in the device by setting which arms are under tension, floating, or grounded.

\section{\label{sec:results-rxy-pulse-amplitude-const-pulse-length} Transverse resistance as a function of pulse amplitude}

In most of the reported current-induced switching experiments, the antiferromagnetic materials were epitaxial films. We thus first characterize the response of the epitaxial samples Al$_2$O$_3$/NiO/Pt and Al$_2$O$_3$/Pt. Following the pulse-sense scheme described in Sect.~\ref{sec:pulsing-and-reading-scheme}, Fig.~\ref{fig: NiO-Pt-comparison}(b)-(c) illustrates the typical variation of the transverse resistance of Al$_2$O$_3$/NiO/Pt and Al$_2$O$_3$/Pt 10~$\mu$m-wide single Hall crosses following the application of orthogonal pulses P1 and P2. The pulse amplitude is gradually increased in steps of 0.25~V. For each step, a sequence of 3 repeats, each made of 5 consecutive P1 pulses followed by 5 consecutive P2 pulses of length $\tau = 1$~ms, is implemented (delimited by the dashed lines). The transverse resistance measured after each pulse is plotted as a blue or red dot for P1 and P2 pulses, respectively. We observe a consecutive decrease and increase of the transverse resistance following P1 and P2 pulses, respectively. The resistance variation increases with the pulse amplitude in both the magnetic and nonmagnetic samples. A change of resistance of $\Delta {R_{xy}} = 0.05 $~$\Omega$ is measured for a pulse amplitude of 9.25~V in Al$_2$O$_3$/NiO/Pt and 11~V in Al$_2$O$_3$/Pt, corresponding to a current density ${j_w}$ along the diagonal of $0.75 \cdot 10^{12}$~A/m$^2$ and $0.76 \cdot 10^{12}$~A/m$^2$, respectively.

The shape of the signal evolves as the devices undergo several repeats. At first, ${R_{xy}}$ has a saw-tooth shape as the resistance changes gradually after each of the 5 consecutive pulses along P1 and P2. At larger voltage, the resistance changes in a step-like manner, where the effect of the first pulse is comparably much larger than the 4 successive pulses [see insets in Fig.~\ref{fig: NiO-Pt-comparison}(b)-(c)]. The saw-tooth shape has been previously associated with resistive changes of the nonmagnetic layer, whereas the step-like shape has been considered as a signature of AFM switching \cite{A003-Chen2018, A004-Moriyama2018, A070-Baldrati2019, A317-Schreiber2020}. This distinction is not supported by the data presented in Fig.~\ref{fig: NiO-Pt-comparison}, in agreement with recent works on magnetic and nonmagnetic samples \cite{A072-Chiang2019, A069-Wagner2019, A138-Churikova2020}. We also observe a stabilization of the amplitude of the resistance changes after 5 consecutive pulses with increasing pulse amplitude, which we assign to a training effect. As the behavior of the Al$_2$O$_3$/NiO/Pt and Al$_2$O$_3$/Pt samples is both qualitatively and quantitatively similar, we attribute the variations of ${R_{xy}}$ to pulse-induced changes of the Pt resistivity.

The variation of ${R_{xy}}$ is attributed to the higher current density passing along the corners of the Hall cross [shaded areas in Fig.~\ref{fig: NiO-Pt-comparison}(a)]. According to the Wheatstone bridge model described in Sect.\ref{sec:wheatstone-bridge}, the negative and positive $\Delta {R_{xy}}$ after pulsing along P1 and P2, respectively, correspond to a decrease of the local resistivity. These observations indicate that the temperature rise during the pulse anneals the metal layer, which decreases its resistivity after the cooling of the device. As the pulse amplitude increases, the annealing process becomes less effective and $\Delta {R_{xy}}$ tends to saturate. This conclusion is corroborated by the work of T. Wagner \textit{et al.} \cite{A069-Wagner2019}, in which Nb thin films deposited at room-temperature on an MgO substrate showed a similar decrease of the resistivity along the pulsed-corners of the device. Furthermore, this behavior agrees with previous studies of the resistivity of Pt films, which show an initial decrease followed by an increase of the Pt resistivity as a function of annealing temperature owing to thermally-induced redistribution of structural imperfections \cite{A149-Xiao2013, A193-Schmid2008}. Finally, measurements performed on the SiO$_2$/NiO/Pt, SiO$_2$/Pt, Si$_3$N$_4$/NiO/Pt and Si$_3$N$_4$/Pt all show a similar behavior with no systematic differences between them that could evidence a magnetic signal.

In the literature, the nonmagnetic variation of ${R_{xy}}$ in AFM/Pt bilayers has been assigned to two other effects. One is the temporary increase of the local resistivity due to Joule heating that persists during the sensing \cite{A138-Churikova2020}. The other is electromigration occurring at the corners of the Hall cross, which causes the formation of hillocks and metal voids that increase the local resistivity \cite{A138-Churikova2020, A069-Wagner2019}. Both these effects would result in $\Delta {R_{xy}}$ of opposite sign with respect to our measurements. Concerning the first effect, the waiting time between the writing and reading events is about 2 s, which is long enough for the device to thermalize to the substrate temperature \cite{A040-You2006, A155-Yamaguchi2005}. As for electromigration, we cannot exclude that it coexists with thermal annealing. Likely, however, electromigration dominates at voltages higher than 20~V in our 10~$\mu$m-wide Hall crosses (${j_w} > 1.7 \cdot 10^{12}$~A/m$^2$), as we also observe an increase of the resistance preceding the device breakdown at high current density (see Sect.~\ref{sec:results-size-effects}). Other models have linked $\Delta {R_{xy}}$ to a Seebeck voltage along the direction of the current-induced temperature gradient in Pt \cite{A072-Chiang2019}. Accordingly, if the Seebeck voltage is at the origin of the change of the transverse resistance, the measured voltage should be independent of the sensing current. Contrary to this hypothesis, we find a linear relationship between $\Delta {V_{xy}}$ and the sensing current between 0.1~mA and 10~mA, which confirms the resistive origin of $\Delta {R_{xy}}$.

\section{\label{sec:results-temporal-relaxation} Temporal relaxation of the transverse resistance after pulsing}

\begin{figure}[]
 \centering
 \includegraphics[width=0.4\linewidth]{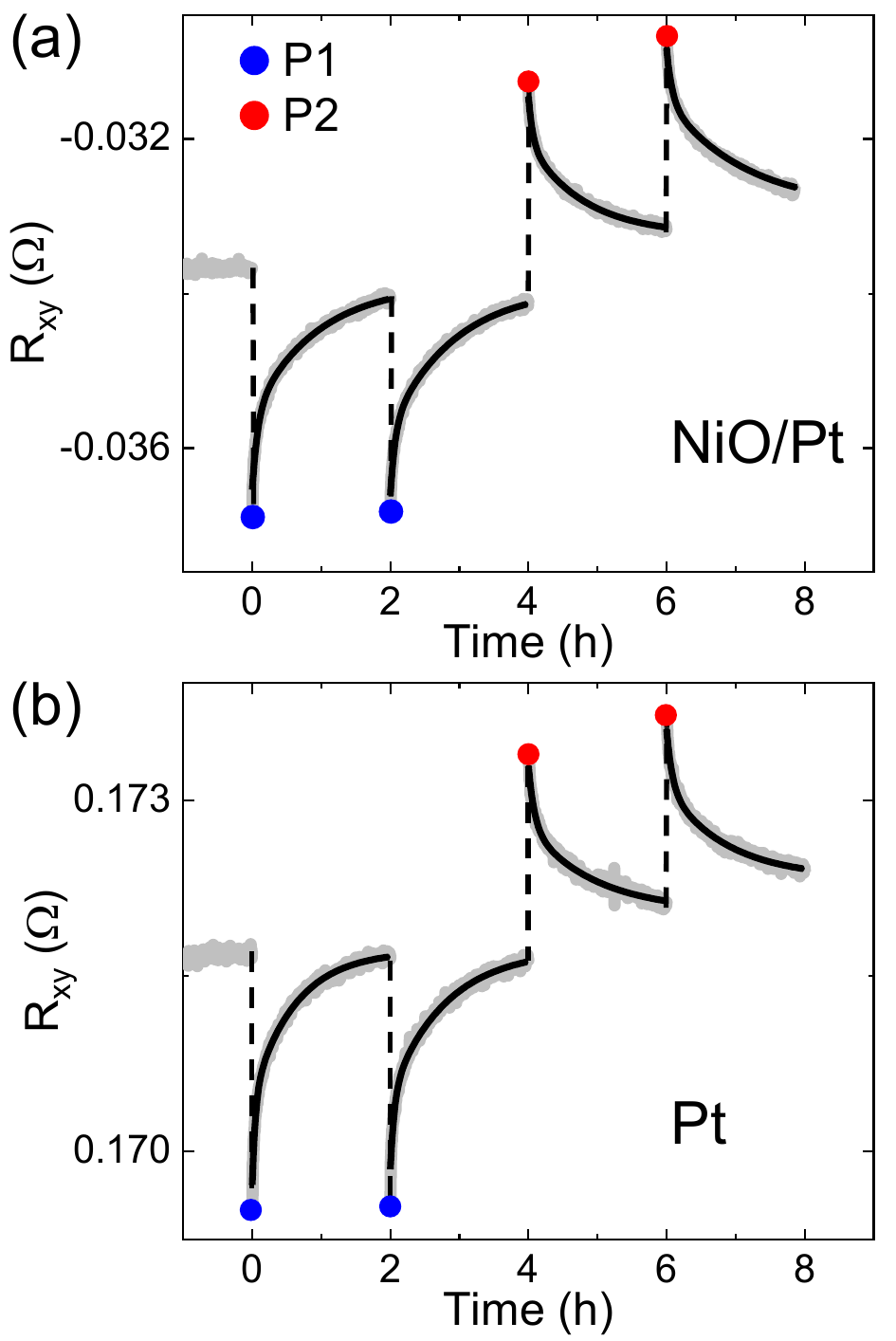}
 \caption{\footnotesize Relaxation of the transverse resistance following single pulses P1 (blue dots) and P2 (red dots) for Al$_2$O$_3$/NiO/Pt (a) and Al$_2$O$_3$/Pt (b) in 10-$\mu$m-wide Hall crosses. The timing of each pulse is indicated by a vertical dashed line. The solid black lines are fits with a double exponential function (see text for details).}
 \label{fig: relaxation}
\end{figure}

After pulsing, ${R_{xy}}$ relaxes towards the pre-pulse value on a time scale of hours. Figure~\ref{fig: relaxation} illustrates the typical evolution of ${R_{xy}}$ over 2~h following single pulses P1 (blue) and P2 (red) in Al$_2$O$_3$/NiO/Pt and Al$_2$O$_3$/Pt. The timing of the single pulses is denoted by the dashed lines. As a reference, ${R_{xy}}$ of the pristine device is recorded starting 1~h before the first pulse P1 to demonstrate the absence of drift in the measurements. The relaxation, which we observe in nonmagnetic and magnetic samples alike, is characterized by different short and long time scales that cannot be captured by a simple exponential fit. To fit the data, we thus use a double exponential function ${R_{xy}}(t) = y_0 + A_{1} \cdot exp(-(t-t_0)/\lambda_{1}) +A_{2} \cdot exp(-(t-t_0)/\lambda_{2})$, where $t_0 = 2$~s is the time interval between the pulse and the sense current, $A_{1,2}$ are amplitude parameters, and $\lambda_{1,2}$ are the relaxation times. Such double exponential decay is typical of relaxation processes in metallic glasses, where the fast process is local and reversible and the slow one involves large scale irreversible structural rearrangements due to the subdiffusive motion of atoms \cite{A209-Luo2015, A056-Luo2017}. 

The fits of P1 and P2 curves (black lines in Fig.~\ref{fig: relaxation}) give a short decay time $\lambda_1 = 4 \pm 0.6$ min and $4.2 \pm 0.2$ min for Al$_2$O$_3$/NiO/Pt and Al$_2$O$_3$/Pt, respectively, and a long decay time $\lambda_2 = 48 \pm 5$ min and $56 \pm 3$ min. These short and long decay times are of the same order of magnitude in the six different samples considered in this study. Also, we find that the fit coefficients do not depend significantly on the substrate or the presence of the NiO layer. A similar relaxation process was observed in MgO/Pt, Si/Pt and glass/Pt\cite{A072-Chiang2019}, but was attributed to slow cooling through the substrate and a long lasting transverse voltage due to the Seebeck effect across the device. This interpretation, however, requires the amplitude of the transverse voltage to be independent of the sensing current, which is not compatible with the linear relation between the transverse voltage and the sensing current found in our measurements (see Section~\ref{sec:results-rxy-pulse-amplitude-const-pulse-length}). Other studies of Pt and NiO/Pt have related the long lasting relaxation of ${R_{xy}}$ to the diffusional creep induced by electromigration with the Pt grain boundaries acting as sources and sinks for point defects \cite{A138-Churikova2020}.

We further note that the long lasting relaxation process is not observed in the consecutive pulse-sense-pulse measurements reported in Fig.~\ref{fig: NiO-Pt-comparison}(b)-(c) as the time interval between these steps is too short (about 2~s). However, in such consecutive measurements, the relaxation manifests itself as a drift of the resistance baseline. Moreover, the relaxation can induce a sign inversion of $\Delta {R_{xy}}$ in consecutive pulse-sense-pulse measurements, which we discuss in the following section.

\section{\label{sec:results-rxy-pulse-amplitude-pulse-length} Transverse resistance as a function of pulse amplitude and pulse length}

\begin{figure*}[]
 \centering
 \includegraphics[width=0.99\linewidth]{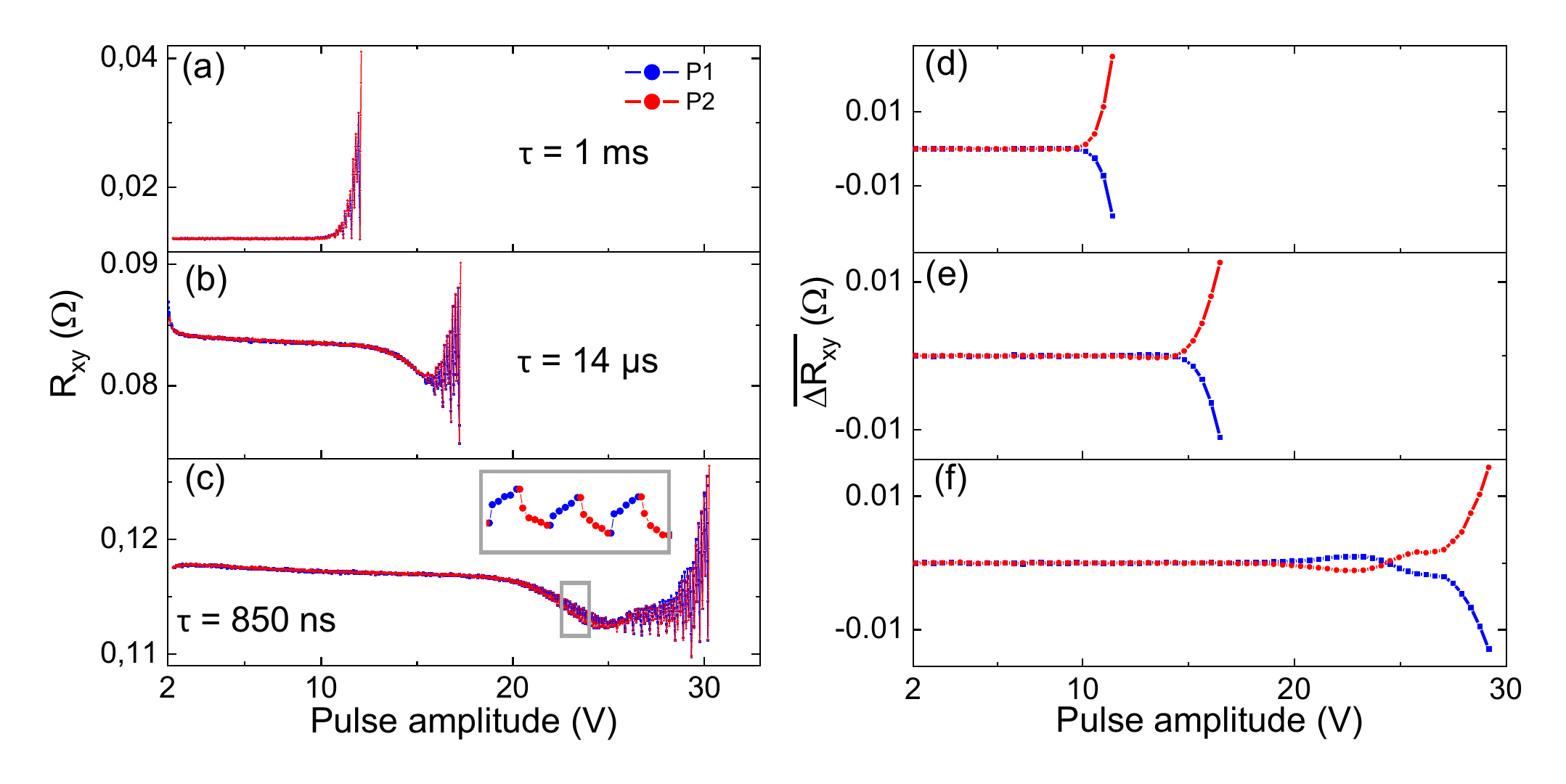}
 \caption{\footnotesize (a)-(c) ${R_{xy}}$ as a function of writing pulse amplitude measured in a Al$_2$O$_3$/Pt 10~$\mu$m-wide Hall cross for pulses of length (a) 1~ms, (b) 14~$\mu$s, and (c) 850~ns. The pulse amplitude is increased in steps of 0.25 V. For each step, a sequence of 3 repeats of 5 consecutive P1 pulses and 5 consecutive P2 pulses is applied, as in Fig.~\ref{fig: NiO-Pt-comparison}(b)-(c). (d)-(f) $\overline{\Delta {R_{xy}}}$ is the average change of resistance after 5 pulses calculated over 3 repeats at constant pulse amplitude from the traces in (a)-(c). The pulsing ramps were stopped when $\overline{\Delta {R_{xy}}} \geq 0.01$~$\Omega$. The inset shows ${R_{xy}}$ trace with inverted sign change.
}
 \label{fig: AOc-traces}
\end{figure*}

In this section, we analyze the response of the devices when the length of the writing pulses is reduced from 1~ms to below 1~$\mu$s. Following the same procedure described in Sect.~\ref{sec:results-rxy-pulse-amplitude-const-pulse-length}, Fig.~\ref{fig: AOc-traces}(a)-(c) shows the change of ${R_{xy}}$ in a Al$_2$O$_3$/Pt 10~$\mu$m-wide Hall cross as a function of pulse amplitude for pulse lengths of 1~ms, 14~$\mu$s, and 850~ns. In order to avoid extensive training and break down of the devices during the measurements, the pulse amplitude is increased in steps of 0.25~V until $\overline{\Delta {R_{xy}}} \geq 0.01 $~$\Omega$, where $\overline{\Delta {R_{xy}}}$ is the arithmetic mean of the change of resistance after 5 pulses over the 3 repeats that are recorded at the same pulse amplitude. Figure~\ref{fig: AOc-traces}(d)-(f) reports $\overline{\Delta {R_{xy}}}$ calculated from the traces in Fig.~\ref{fig: AOc-traces}(a)-(c). We observe that the minimum pulse amplitude required to obtain a noticeable $\overline{\Delta {R_{xy}}} \geq 0.001 $~$\Omega$ increases systematically for shorter pulses: 10.5~V, 11~V, and 20~V for a pulse length of 1~ms, 14~$\mu$s, and 850~ns, respectively. This highly nonlinear relation between the pulse amplitude and pulse length suggests an interplay between the current-induced Joule heating of the device and heat dissipation through the substrate. Such behavior is expected if $\overline{\Delta {R_{xy}}}$ is due to current-induced annealing, as the probability of thermally activated processes depends linearly on time and exponentially on the inverse of the temperature.

The curves shown in Fig.~\ref{fig: AOc-traces} evidence also a variation of the baseline of ${R_{xy}}$, which changes from straight in (a) to nonmonotonic in (b) and (c). This effect is ascribed to the training of the device. In a pristine device, as the voltage is ramped up for the first time, ${R_{xy}}$ decreases after P1 pulses and comes back to its previous value after an equal number of P2 pulses, as shown in Fig.~\ref{fig: AOc-traces}(a). As the ramp in Fig.~\ref{fig: AOc-traces}(a) ends with 5 pulses along P2, ${R_{xy}}$ is large and tends to relax at the beginning of the second ramp, shown in Fig.~\ref{fig: AOc-traces}(b). Depending on the length, spacing, and amplitude of the pulses, the interplay of excitation and relaxation effects results in a changing baseline of ${R_{xy}}$, as discussed in more detail in Sect.~\ref{sec:results-training-effects}. 

\begin{figure*}
 \centering
 \includegraphics[width=0.99\linewidth]{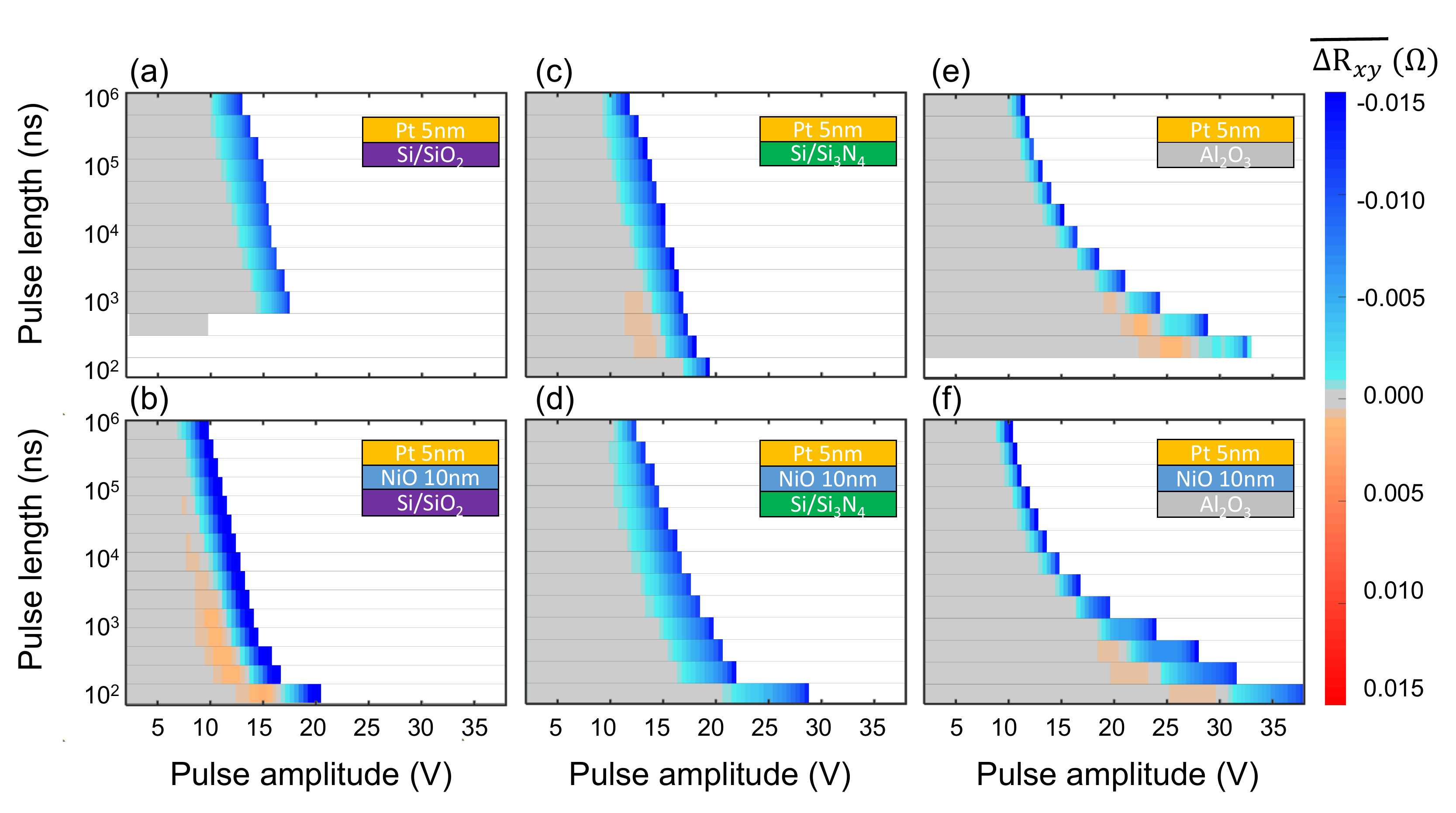}
 \caption{\footnotesize Comparison of the average transverse resistance change as a function of pulse amplitude and pulse length in a 10-$\mu$m-wide Hall cross of (a) SiO$_2$/Pt, (b) SiO$_2$/NiO/Pt, (c) Si$_3$N$_4$/Pt, (d) Si$_3$N$_4$/NiO/Pt, (e) Al$_2$O$_3$/Pt and (f) Al$_2$O$_3$/NiO/Pt. The plots show $\overline{\Delta {R_{xy}}}$ measured after 5 consecutive P1 pulses and averaged over 3 repeats at constant pulse amplitude, following the same measurement scheme reported in Fig.\ref{fig: AOc-traces}. For each pulse length, a ramp of increasing pulse amplitude is applied until $\overline{\Delta R_{xy}}\geq 0.01 \Omega$. The color scale represents the amplitude and sign of $\overline{\Delta R_{xy}}$. Grey-shaded areas indicate no resistance change.}
 \label{fig: substrate-all-devices}
\end{figure*}

Measurements performed with shorter pulses evidence another remarkable effect, namely the inversion of the sign of $\Delta {R_{xy}}$ preceding the onset of the larger variations of $\Delta {R_{xy}}$, which can be observed in the inset of Fig.~\ref{fig: AOc-traces}(c). The sign inversion becomes more clear when plotting $\overline{\Delta {R_{xy}}}$, namely the average excursion of the trace as a function of the pulse amplitude. At high voltage, $\overline{\Delta {R_{xy}}}$ is consistent with a decrease of ${R_{xy}}$ induced by pulsing, as described in Sect.~\ref{sec:results-rxy-pulse-amplitude-const-pulse-length}. At intermediate amplitude, however, the sign of $\overline{\Delta {R_{xy}}}$ is inverted, as seen by the difference of the blue and red curves in Fig.~\ref{fig: AOc-traces}(f) and in the inset of Fig.~\ref{fig: AOc-traces}(c). The sign inversion typically appears when multiple pulses are applied to Pt and NiO/Pt devices and when the training state is more advanced.

We assign this transient sign inversion to heat-assisted relaxation induced by a current pulse, which can dominate over annealing in trained devices. During the first ramp, each pair of the device quadrants, ${R_{1,4}}$ and ${R_{2,3}}$ (see Sect.~\ref{sec:wheatstone-bridge}), undergoes the same annealing effect and $\Delta {R_{xy}}$ for P1 (P2) is successively negative (positive). Then, in the small amplitude regime at the beginning of the following ramps, the four quadrants uniformly relax and give $\Delta {R_{xy}}\approx 0$. Upon increasing the pulse amplitude, the temperature of the quadrants gradually rises, which speeds up the relaxation of these quadrants compared to the unpulsed quadrants. This asymmetric change of resistance induces a positive (negative) $\Delta {R_{xy}}$ after P1 (P2) pulses. This transient state takes place only in a small range of pulse amplitudes, namely when the current supplies enough heat to accelerate the relaxation rate without initiating the annealing process. The amplitude of this effect depends on the training history of the sample and changes from device to device.

\section{\label{sec:results-influence-substrate} Influence of the antiferromagnetic layer and substrate}

In order to investigate the influence of the NiO layer and of the substrate on the transverse resistance, we compare the average $\overline{\Delta {R_{xy}}}$ as a function of the pulse amplitude and pulse length for Pt and NiO/Pt deposited on Si/SiO$_2$, Si/Si$_3$N$_4$ and Al$_2$O$_3$, as shown in Fig.~\ref{fig: substrate-all-devices}(a)-(f). In these plots, the colors represent the amplitude of $\overline{\Delta {R_{xy}}}$ measured after P1 pulses and the grey areas represent the range where $\overline{\Delta {R_{xy}}}\approx 0$. No measurements were acquired in the white areas because the change of resistance exceeded the limit $\overline{\Delta {R_{xy}}} \geq 0.01 $~$\Omega$. The measurements of $\overline{\Delta {R_{xy}}}$ recorded after P2 pulses are roughly symmetric and opposite in sign to P1, corresponding to complementary color maps with inverted blue and red colors.

We find that the presence of NiO makes very little difference in the response of the devices. Neither the NiO/Pt sample with high epitaxial quality grown on Al$_2$O$_3$ nor those grown on SiO$_2$ and Si$_3$N$_4$ show significant differences with respect to the single Pt layers, confirming that the resistance variation in such structures originates from Pt rather than from switching of the AFM. 

In all devices, the pulse amplitude required to attain a given variation of ${R_{xy}}$ increases as the pulse length is reduced. However, the rate at which $\overline{\Delta {R_{xy}}}$ rises as a function of pulse amplitude and pulse length is closely related to the substrate. In order to maintain the same $\overline{\Delta {R_{xy}}}$ amplitude when reducing the pulse length by a factor 1000, the pulse amplitude needs to be increased by a factor 2 for the samples grown on SiO$_2$ and Si$_3$N$_4$ and by a factor of 3 for the samples grown on Al$_2$O$_3$. This observation can be rationalized in terms of the thermal diffusivity $\mu_s = \kappa_s/(\rho_s \, c_s) $ of the substrate, which exemplifies how well a material conducts heat in terms of the thermal conductivity $\kappa_s$ relative to its volumetric capacity for storing thermal energy, as given by the product of the density $\rho_s$ and specific heat $c_s$. At room-temperature, the thermal diffusivity calculated from thin film parameters is $\mu_\text{SiO2}\approx 0.8$~mm$^2\,$s$^{-1}$ for SiO$_2$ ($\kappa = 1.3$~W$\,$m$^{-1}\,$K$^{-1}$, $\rho = 2200$~kg$\,$m$^{-3}$, $c = 730$~J$\,$kg$^{-1}\,$K$^{-1}$), $\mu_\text{Si3N4}\approx 0.9$~mm$^2\,$s$^{-1}$ for Si$_3$N$_4$ ($\kappa = 1.4$~W$\,$m$^{-1}\,$K$^{-1}$, $\rho = 2100$~kg$\,$m$^{-3}$, $c = 730$~J$\,$kg$^{-1}\,$K$^{-1}$~ \cite{A040-You2006, A167-Lee1997}), and from bulk parameters $\mu_\text{Al2O3}\approx 28$~mm$^2\,$s$^{-1}$ for Al$_2$O$_3$ ($\kappa = 46.06 $~W$\,$m$^{-1}\,$K$^{-1}$, $\rho = 3980$~kg$\,$m$^{-3}$, $c = 418$~J$\,$kg$^{-1}\,$K$^{-1}$~ \cite{z-sapphire-sample-provider}). The gray shaded areas in Fig.~\ref{fig: substrate-all-devices} show that the substrates with high thermal diffusivity allow for using a larger range of pulse amplitudes and pulse lengths without inducing changes of ${R_{xy}}$. This finding is explained by the fact that substrates with a large thermal diffusivity can transfer heat fast without heating up too much compared to substrates with a low thermal diffusivity. 

As a corollary, we note that the heat effectively dissipated in Pt depends on the balance between electrical power supplied to the system, heat transfer by conduction through the NiO and the substrate, and convection and radiation at the top of the devices. However, the convective and radiation heat losses are much smaller than the conduction heat transfer \cite{A240-Stojanovic2007} and roughly equal for all the devices investigated in this study. Therefore, we conclude that the thermal diffusivity of the substrate is the main parameter affecting the heat accumulation in Pt induced by a given current in devices of equal size.

\begin{figure}
 \centering
 \includegraphics[width=0.5\linewidth]{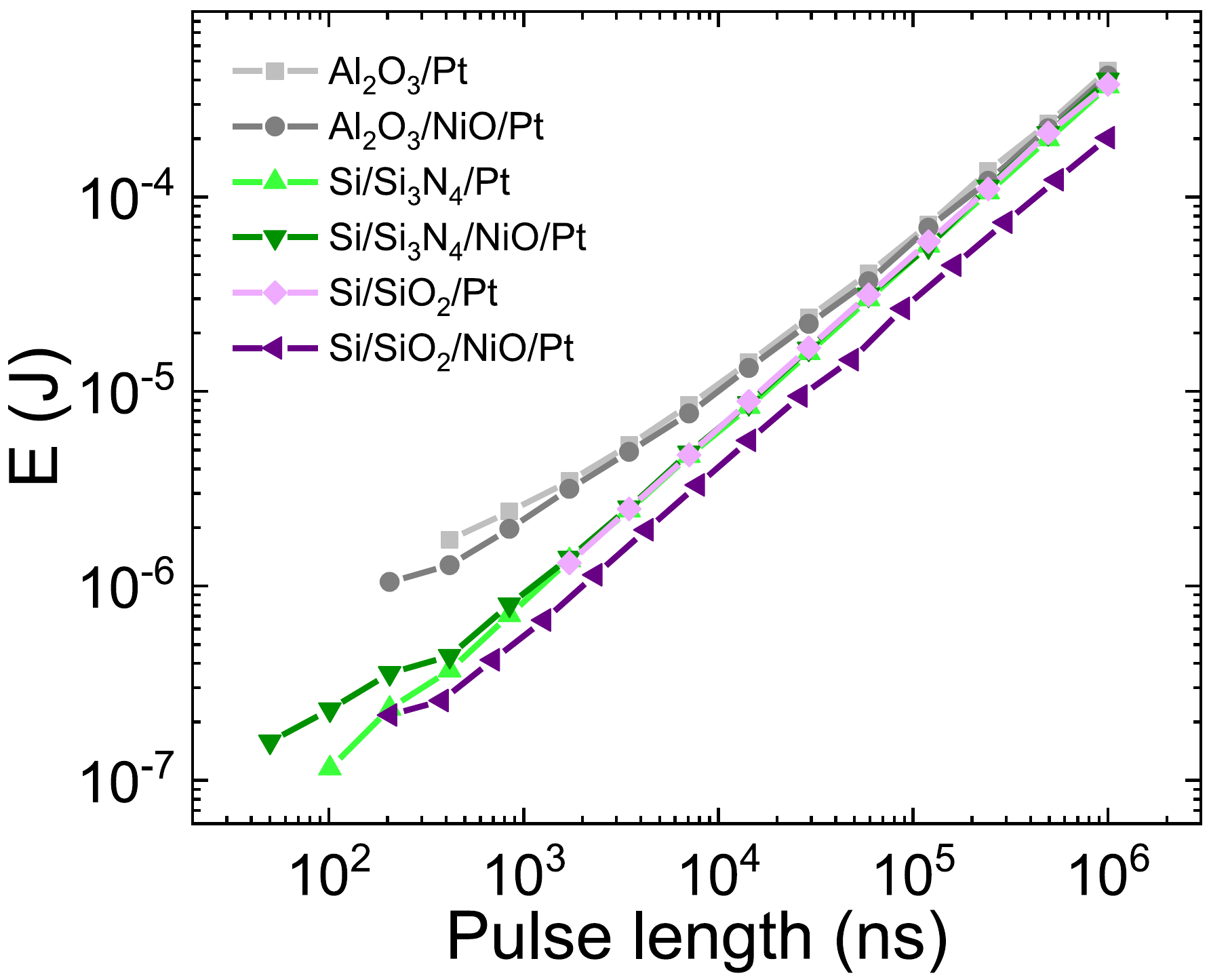}
 \caption{\footnotesize Electrical energy ${E = \tau V^2/R_{p}}$ required to obtain a relative change of the transverse resistance $\overline{\Delta {R_{xy}}}/\Delta {R_{xx}} \geq 10^{-5}$ as a function of pulse length computed from the data in Fig.~\ref{fig: substrate-all-devices}.} \label{fig: pulse-length-energy}
\end{figure}

The temperature increase attained during the pulse is related to the electrical energy dissipated in the device, $E = \tau V^2 / R_p$, where ${\tau}$ is the pulse length, ${R_p}$ the resistance of the pulse path, and ${V}$ the applied voltage. Figure~\ref{fig: pulse-length-energy} plots the energy required to induce a relative change $\overline{\Delta {R_{xy}}}/{R_{xx}} \geq 10^{-5}$ as a function of pulse length. The curves confirm that samples grown on SiO$_2$ and Si$_3$N$_4$ substrates require less energy compared to Al$_2$O$_3$ in order to attain a similar change in resistance. Moreover, when comparing the Al$_2$O$_3$ and Si-based substrates, the curves follow the same trend for long pulses but diverge for short pulses, showing that the samples grown on Al$_2$O$_3$ dissipate heat more efficiently for shorter pulses compared to longer pulses. We attribute this trend to the timescale required to transfer heat from the device to the substrate. For short pulses, $\tau \leq 10$~$\mu$s, the temperature rises sharply in the device but not in the substrate. For longer pulses, on the other hand, the heat dissipated in the device diffuses to the substrate and warms it up. The heat diffusivity of the Si-based substrates varies little with increasing temperature \cite{A167-Lee1997, A236-Griffin1994}, whereas the heat diffusivity of Al$_2$O$_3$ drops from $\sim$28~~mm$^2\,$s$^{-1}$ at room temperature, to $\sim$15~~mm$^2\,$s$^{-1}$ at 100~$^\circ$C, and $\sim 8$~~mm$^2\,$s$^{-1}$ at 400~$^\circ$C~ \cite{z-sapphire-sample-provider}. Therefore, Al$_2$O$_3$ becomes significantly less efficient in dissipating heat for the longer pulses, which explains the converging trend of the curves shown in Figure~\ref{fig: pulse-length-energy}.

\section{\label{sec:results-size-effects} Size effects}

\begin{figure}
 \centering
 \includegraphics[width= 0.5\linewidth]{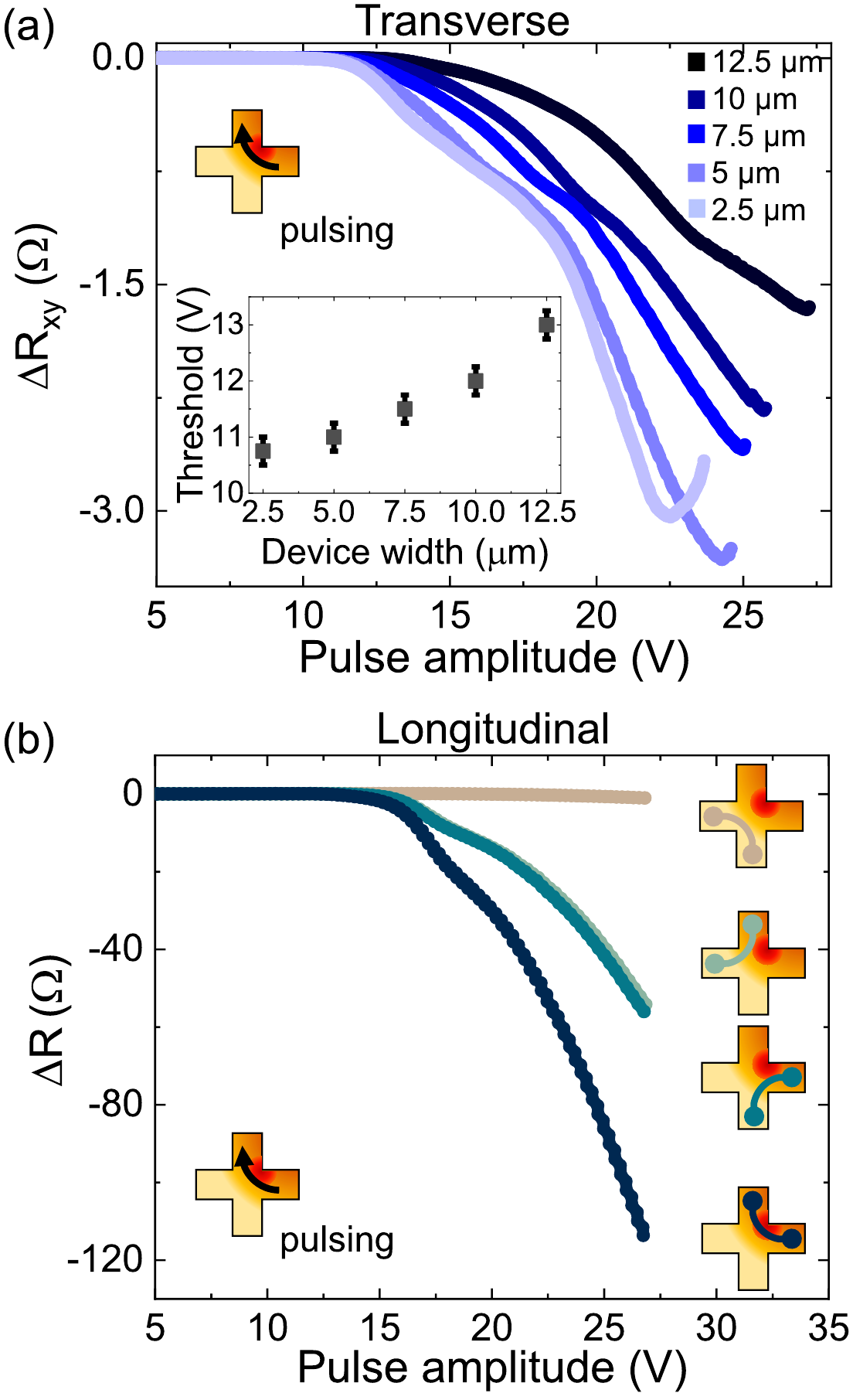}
 \caption{\footnotesize (a) Change of the transverse resistance $\Delta {R_{xy}}$ as a function of pulse amplitude for pulses sent around one corner of a Hall cross (see arrow). The data are measured in Si$_3$N$_4$/Pt(5nm) Hall crosses of different widths. (b) Change of the longitudinal resistance ${\Delta R}$ of the four quadrant paths of a 5-$\mu$m-wide cross as a function of pulse amplitude for pulses sent around one corner of the cross. The pulsing and measuring geometry are indicated in the schematic diagrams. The initial resistance of each path is, on average, ${R_0} = 703\pm 1$~$\Omega$.}
 \label{fig: size-effect}
\end{figure}

The power density dissipated in Pt is given by ${V^2/(\rho \, L^2)}$, where ${L}$ is the effective length of the pulse path, which is approximately independent of the device width. As the current distribution in the Hall cross is inhomogeneous, however, we find that the device width strongly influences the amplitude of $\Delta {R_{xy}}$. More precisely, we observe that $\Delta {R_{xy}}$ increases as the ratio between the area of the Hall cross and the length of the excited path decreases. Figure~\ref{fig: size-effect}(a) reports $\Delta {R_{xy}}$ as a function of pulse amplitude for Si$_3$N$_4$/Pt Hall crosses with width ranging from 2.5~$\mu$m to 12.5~$\mu$m in steps of 2.5~$\mu$m. To study the asymmetry induced by the writing current, only the top right quadrant of the device is pulsed, exciting mainly ${R_1}$ (see Sect.~\ref{sec:wheatstone-bridge}), in sequences of 10 pulse-repeats starting from 2~V and increasing in steps of 0.25~V until the device fails.

We observe that $\Delta {R_{xy}}$ is negative and decreases markedly up to a pulse amplitude of about 25~V, after which it starts increasing just before reaching the device breakdown. As argued in Sect.~\ref{sec:results-rxy-pulse-amplitude-const-pulse-length}, the decrease of ${R_{xy}}$ in this geometry corresponds to a decrease of the local resistivity, whereas the final upturn corresponds to an increase of the local resistivity. These measurements support the conclusion that competing effects take place in the crosses, consistently with thermal annealing and electromigration dominating the intermediate and higher voltage range, respectively. Importantly, smaller crosses present larger $\Delta {R_{xy}}$ signals compared to larger crosses. The inset in Fig.~\ref{fig: size-effect}(a) shows that the pulse amplitude threshold for observing $\Delta {R_{xy}} \geq 0.01 $~$\Omega$ increases with the width of the cross, which we attribute to the larger proportion between the device area and the current-crowded area.

The variation of the longitudinal resistance $\Delta R$ measured between two adjacent arms of the cross, shown in Fig.~\ref{fig: size-effect}(b), is consistent with the decrease of ${R_{xy}}$ derived from the Wheatstone model. Moreover, we observe that the resistance changes not only occurs at the corresponding corner where the current density is the highest, but also in its surroundings and the pulsed arms. Before the breakdown of the device, the maximum relative change of the longitudinal resistance is ${\Delta R/R} \approx - 16$~$\%$ measured along the pulse path (black curve), $\approx - 8$~$\%$ in the two adjacent quadrant paths (light and dark green curves), and $\approx - 0.14$~$\%$ in the quadrant opposite to the pulse path (orange curve). 

\section{\label{sec:results-training-effects} Influence of training effects on the resistance baseline}

\begin{figure}
 \centering
 \includegraphics[width=\linewidth]{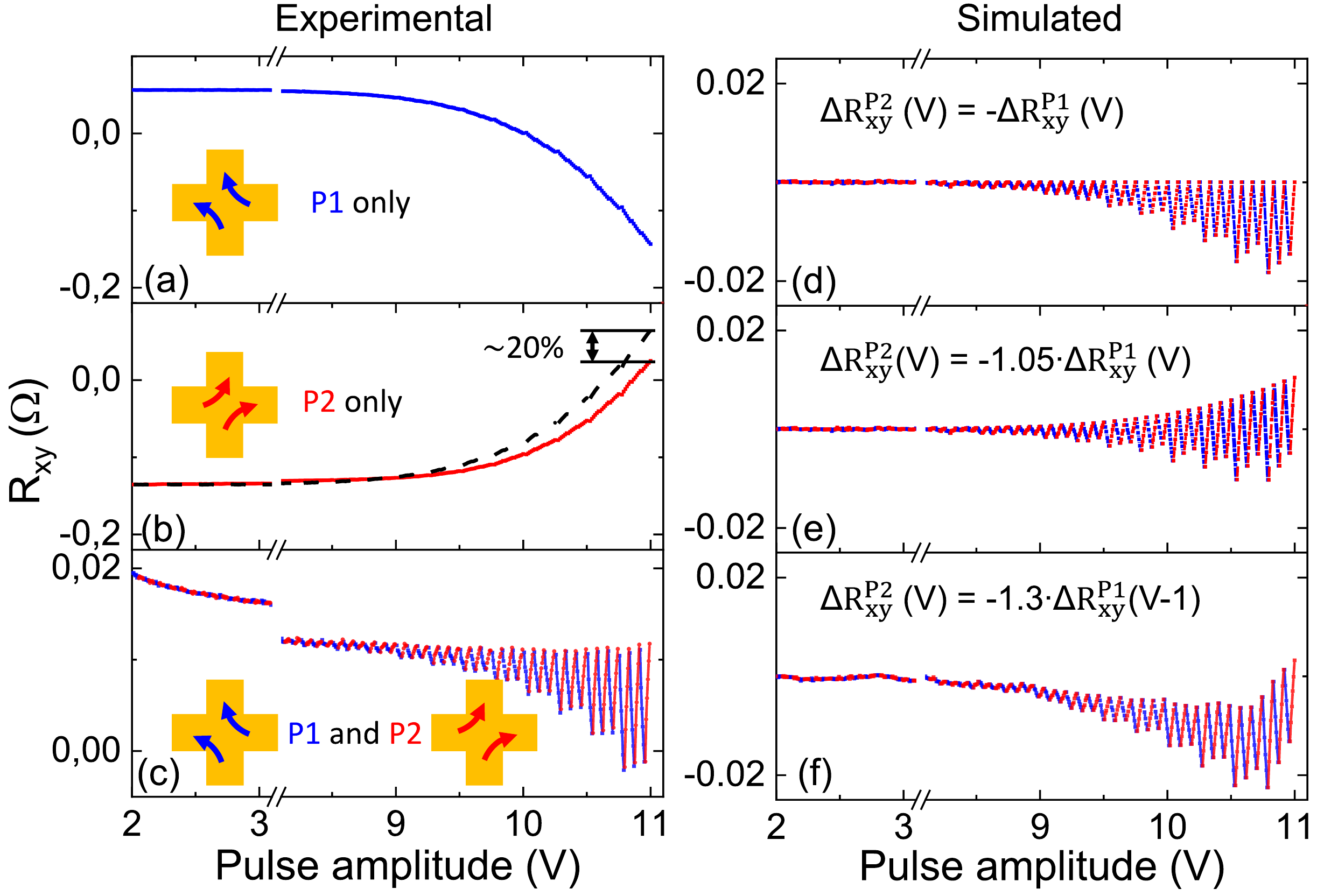}
 \caption{\footnotesize Variation of ${R_{xy}}$ in a 10-$\mu$m-wide Si$_3$N$_4$/Pt cross as a function of pulse amplitude for P1 (a), P2 (b), and P1 and P2 pulses (c). In each ramp, the pulse amplitude is increased in steps of 0.25 V. For each step, a sequence of 3 repeats of 5 consecutive P1 or P2 pulses of length $\tau = 1$~ms is applied. (d) Simulated device response to P1 and P2 pulsing calculated from the curves shown in (a,b) by assuming a symmetric response along P1 and P2, $[\Delta R^{P2}_{xy}(V) = -\Delta R^{P1}_{xy}(V) ]$, (e) a $5\%$ higher $\Delta {R_{xy}}$ for P2 pulses, $[\Delta R^{P2}_{xy}(V) = -1.05\cdot\Delta R^{P1}_{xy}(V)]$, and (f) a $30\%$ higher amplitude and a 1~V shift of the threshold for P2 pulses, $[\Delta R^{P2}_{xy}(V) = -1.3\cdot\Delta R^{P1}_{xy}(V-1)]$.
}
 \label{fig: empirical}
\end{figure}

In this section, we analyze the nonlinear variation of the baseline of ${R_{xy}}$ as a function of the pulse amplitude, which determines the shape of the traces shown in Fig.~\ref{fig: NiO-Pt-comparison}(b)-(c) and Fig.~\ref{fig: AOc-traces}(a)-(c). We argue that this shape is due to training of the devices. To understand the effects of training, we first measure ${R_{xy}}$ after pulsing the device only along P1, as shown in Fig.~\ref{fig: empirical}(a), then only along P2, as shown in Fig.~\ref{fig: empirical}(b), and finally run the sequence of alternate P1 and P2 pulses already described in Sect.~\ref{sec:results-rxy-pulse-amplitude-const-pulse-length}, as shown in Fig.~\ref{fig: empirical}(c). The ramps shown in Fig.~\ref{fig: empirical} (a)-(c) are taken consecutively with no waiting time between each other. During the first ramp, ${R_{xy}}$ progressively decreases with the increasing number of P1 pulses. The signal of the second ramp starts from the value of ${R_{xy}}$ attained at the end of the first ramp and varies in the opposite direction with an increasing number of P2 pulses. We note that here the relaxation process (Fig.~\ref{fig: relaxation}) is not visible because of the short period of time elapsed during the ramp. By overlapping the first and second ramp [black dashed line and red line in Fig.~\ref{fig: empirical}(b)], we observe that the end value of ${R_{xy}}$ in the P2 ramp is smaller by about $\approx 20$~$\%$ compared to the P1 ramp. This asymmetry between the P1 and P2 response is attributed to the training of the device during the first ramp, as P1 pulses also partially affect the resistance of the P2 pulse path (see Fig.~\ref{fig: size-effect}). This interpretation was confirmed by inverting the order of the pulses, namely when the P2 ramp preceded the P1 ramp in a pristine device, which led to the opposite asymmetry. Thus, the amplitude of $\Delta {R_{xy}}$ following P1 and P2 pulses depends on the pulsing history as well as from possible structural imperfections in the Hall cross due to inhomogeneous growth or patterning.

When considering a ramp of alternating P1 and P2 pulses, the asymmetry of $\Delta {R_{xy}}$ between P1 and P2 pulses results in a nonmonotonic baseline of ${R_{xy}}$ that can be mistaken for drift, but is actually the signature of unequal responses to P1 and P2 pulses. In Figure~\ref{fig: empirical}(c), as the same number of pulses have been applied to the four corners in (a) and (b), the resistance changes along the paths of P1 and P2 are partially balanced. Hence, the ramp with alternate pulses along P1 and P2 is close to symmetric. In general, however, the device never goes back to a fully balanced state after pulsing along P1 and P2, giving rise to different baseline shapes, as observed, e.g., in Fig.~\ref{fig: NiO-Pt-comparison}(b)-(c) and Fig.~\ref{fig: AOc-traces}(a)-(c). To see how this occurs in practice, we have simulated the P1 and P2 pulsing ramps starting from the curves shown in Fig.~\ref{fig: empirical}(a)-(b), computing the sum of the points measured along P1 and P2 (alternating 5 P1 pulses and 5 P2 pulses), and artificially removing or introducing asymmetries in response to P2 pulses (${\Delta R^{P2}_{xy}(V)}$) relative to P1 (${\Delta R^{P1}_{xy}(V)}$). The results are shown in Fig.~\ref{fig: empirical}(d)-(f), where we reproduce different baselines by assuming (d) a symmetric response with ${\Delta R^{P2}_{xy}(V) = -\Delta R^{P1}_{xy}(V)}$, (e) a $5\%$ higher response along P2, ${\Delta R^{P2}_{xy}(V) = -1.05\cdot\Delta R^{P1}_{xy}(V)}$, and (f) an asymmetric response of the device along P2 by shifting the voltage axis by 1~V and increasing the amplitude of the response by $30 \%$ such that ${R^{P2}_{xy}(V) = -1.3 \cdot \Delta R^{P1}_{xy}(V-1)}$. In agreement with the experiments, we see that the resistance change can take three forms: the baseline of $\Delta {R_{xy}}$ can stay constant at the first value of the series with ${R_{xy}}$ moving back and forth with pulses along P1 and P2 [Fig.~\ref{fig: empirical}(d)], tilt upwards or downwards [Fig.~\ref{fig: empirical}(e)], or describe a nonmonotonic curve similar to a "pipe" [Fig.~\ref{fig: empirical}(f)]. 

Interestingly, the shape of ${R_{xy}}$ reported in Fig.~\ref{fig: empirical}(e) can be misinterpreted as a symmetric response to P1 and P2 pulses because the excursion of ${R_{xy}}$ appears to be symmetric around a central value. However, a symmetric response implies the ability to return always to the initial value of ${R_{xy}}$, as shown in Fig.~\ref{fig: empirical}(d). More complex shapes are obtained when there is an asymmetry in both the amplitude response and threshold voltage required to induce appreciable changes of ${R_{xy}}$, as shown in Fig.~\ref{fig: empirical}(f). Note that in Ref.\cite{A107-Nair2019}, a similar shape as the one reported in Fig.~\ref{fig: empirical}(f) has been associated with a partial reorientation of the Néel vector in an antiferromagnetic transition metal dichalcogenide. However, we repeatedly observed this behavior in Pt samples as well as NiO/Pt bilayers (see Fig.~\ref{fig: empirical}(c)), which shows that special care should be taken when analyzing the signal trace in such a device geometry and pulsing configuration.

\section{\label{sec:conclusion} Conclusions}

In conclusion, we presented a comparative analysis of the current-induced resistance variation in Pt and NiO/Pt samples. A systematic study of the pulse amplitude and pulse length dependence of the transverse resistance performed on epitaxial and non-epitaxial films did not evidence significant differences between Pt and NiO/Pt layers. All our results are consistent with changes of the Pt resistivity occurring at the corners of the Hall crosses due to the current crowding effect. As also seen in recent works \cite{A072-Chiang2019, A069-Wagner2019, A138-Churikova2020}, the resistance can change in a saw tooth or step-like manner in both magnetic and nonmagnetic samples, and therefore the signal shape cannot be used to unambiguously identify magnetic switching, even if ex-situ imaging confirms the reversal of antiferromagnetic domains \cite{A003-Chen2018, A004-Moriyama2018, A070-Baldrati2019, A317-Schreiber2020}. Using a simple Wheatstone bridge model of a Hall cross, we identified competing effects responsible for the resistance changes: a decrease of the local resistivity surrounding the pulsed corner due to thermal annealing of the Pt layer and an increase of the resistivity due to electromigration preceding the device breakdown. After pulsing, the transverse resistance relaxes back towards the value before the pulse following a double exponential law with a short and long relaxation time of about a few minutes and 1 hour, respectively. The interplay between voltage amplitude and relaxation can give rise to nonmonotonic changes and even sign inversion of the variation of resistance. The substrate, apart from determining the crystalline quality of the samples, plays an essential role in dissipating heat away from the Pt line during pulsing. Substrates with a large thermal diffusivity such as sapphire allow for using a significantly broader range of pulse amplitudes without causing changes of the resistance due to current-induced annealing of Pt, thus providing a better chance of measuring magnetic switching by electrical means without artifacts.
At constant pulse amplitude, larger devices show reduced resistance changes compared to smaller devices owing to the larger proportion between the device area and the current-crowded area. Finally, we find that the current-induced changes of the transverse resistance are superimposed onto a nonmonotonic baseline whose shape depends on structural imperfections and device asymmetry as well as on the pulsing history of the sample. Our results provide a systematic overview of current-induced resistance changes in single metal layers and antiferromagnet/metal bilayers. These signals have a nonmagnetic origin and might either obscure or overlap with the magnetoresistive signals due to the switching of antiferromagnetic domains in antiferromagnet/metal bilayers.

\section*{Acknowledgments}
This work was funded by the Swiss National Science Foundation (grant number PZ00P2-179944).


\begingroup
\bibliographystyle{naturemag}
\bibliography{main.bib}
\endgroup

\end{document}